\documentclass[twocolumn,prl,aps,floatfix,superscriptaddress]{revtex4}

\usepackage{graphicx}   
\usepackage{color}      
\usepackage{bm} 
\begin{document}

\title{Superconducting gap structure of BaFe$_2$(As$_{1-x}$P$_{x}$)$_2$}

\author{L. Malone}
\affiliation{H. H. Wills Physics Laboratory, University of Bristol, Tyndall Avenue, Bristol, BS8 1TL, United Kingdom.}

\author{Y. Mizukami}
\affiliation{Department of Physics, Kyoto University, Sakyo-ku, Kyoto 606-8502, Japan.} \affiliation{Department of
Advanced Materials Science, University of Tokyo, Kashiwa, Chiba 277-8561, Japan}

\author{P. Walmsley}
\affiliation{H. H. Wills Physics Laboratory, University of Bristol, Tyndall Avenue, Bristol, BS8 1TL, United Kingdom.}

\author{C. Putzke}
\affiliation{H. H. Wills Physics Laboratory, University of Bristol, Tyndall Avenue, Bristol, BS8 1TL, United Kingdom.}

\author{S. Kasahara}
\affiliation{Department of Physics, Kyoto University, Sakyo-ku, Kyoto 606-8502, Japan.}

\author{T. Shibauchi}
\affiliation{Department of Physics, Kyoto University, Sakyo-ku, Kyoto 606-8502, Japan.} \affiliation{Department of
Advanced Materials Science, University of Tokyo, Kashiwa, Chiba 277-8561, Japan}

\author{Y. Matsuda}
\affiliation{Department of Physics, Kyoto University, Sakyo-ku, Kyoto 606-8502, Japan.}

\author{A. Carrington}
\affiliation{H. H. Wills Physics Laboratory, University of Bristol, Tyndall Avenue, Bristol, BS8 1TL, United Kingdom.}

\begin{abstract}
We present a study of the superconducting gap structure in the iron-pnictide series BaFe$_2$(As$_{1-x}$P$_{x}$)$_2$. By
measuring the variation of the specific heat as a function of temperature and magnetic field we are able to determine
the number and Fermi surface location of the nodes in the superconducting gap. In particular, from measurements of the
variation of the specific heat as the magnetic field is rotated in the $ab$ plane of the sample we conclude that the
nodes are in the [110] directions. Then from a quantitative analysis of the temperature and field dependence of the
specific heat we further conclude that nodes exists on all Fermi surface sheets.
\end{abstract}
\maketitle

Amongst the different families of unconventional superconductors, the iron-pnictides seem to be unique in that the
structure of the superconducting energy gap $\Delta$, i.e., how it varies with momentum $\bm{k}$ on each Fermi surface
sheet, can vary dramatically between different materials \cite{Carrington2011a,Hirshfeld11}. This is despite the fact
that the Fermi surfaces of all the iron-pnictides are quite similar, consisting of quasi-two-dimensional electron and
hole pockets \cite{Carrington2011fs,Kordyuk2012}. It has been proposed theoretically that subtle differences in the
Fermi surface topology, and possible structure of the magnetic interactions, can lead to a switch between different gap
structures \cite{Kuroki09,GraserMHS09,Thomale11}. Also it has been proposed that certain gap structure can only be
explained by orbital fluctuations perhaps in combination with spin fluctuations \cite{shimojima2011orbital,Yoshida14}.
The structure of $\Delta$ and how this varies between different iron-pnictide compounds therefore provides a strong
test to discriminate between candidate microscopic theories of the superconductivity \cite{Hirshfeld11}.

The materials which provide the best discrimination are those which have the most structure in $\Delta(\bm{k})$. In
this sense, BaFe$_2$(As$_{1-x}$P$_{x}$)$_2$ is perhaps the most important example because it has a nodal gap structure
\cite{hashimoto2010line} and also a high $T_c$, reaching a maximum value of 30\,K at $x=0.30$. In most other
iron-pnictides $\Delta(\bm{k})$ does not exhibit nodes \cite{Hirshfeld11} although in some systems the experimental
evidence is somewhat ambiguous because of strong impurity scattering. LaFePO \cite{FletcherPRL09}, LiFeP
\cite{Hashimoto2012} are exceptions which do show evidence for nodes but both of these have a low $T_c$ ($<7$\,K).
BaFe$_2$(As$_{1-x}$P$_{x}$)$_2$ also has the advantage that in high quality single crystals, the electron mean free
path is very high, as evidenced by the observation of de Haas-van Alphen oscillations
\cite{shishido2010evolution,Walmsley2013}, and so the gap structure is not strongly smeared by impurity scattering.

Although bulk probes including magnetic penetration depth $\lambda$ \cite{hashimoto2010line}, nuclear magnetic
resonance \cite{nakai201031} and thermal conductivity \cite{hashimoto2010line,yamashita2011nodal} have shown that
BaFe$_2$(As$_{1-x}$P$_{x}$)$_2$ has a nodal gap structure, the position, number, and orientation of these nodes is
still unclear. Although angle resolved photoemission spectroscopy (ARPES) can in principle resolve this issue results
to date have been inconsistent \cite{shimojima2011orbital,zhang2012nodal,Yoshida14}. Here we show from a study of the
temperature and magnetic field-angle dependence of the specific heat that BaFe$_2$(As$_{1-x}$P$_{x}$)$_2$ has
\textit{vertical} line nodes in the [110] directions and that nodes exist on both the electron and the hole sheets.

\begin{figure}
\includegraphics[width=80mm]{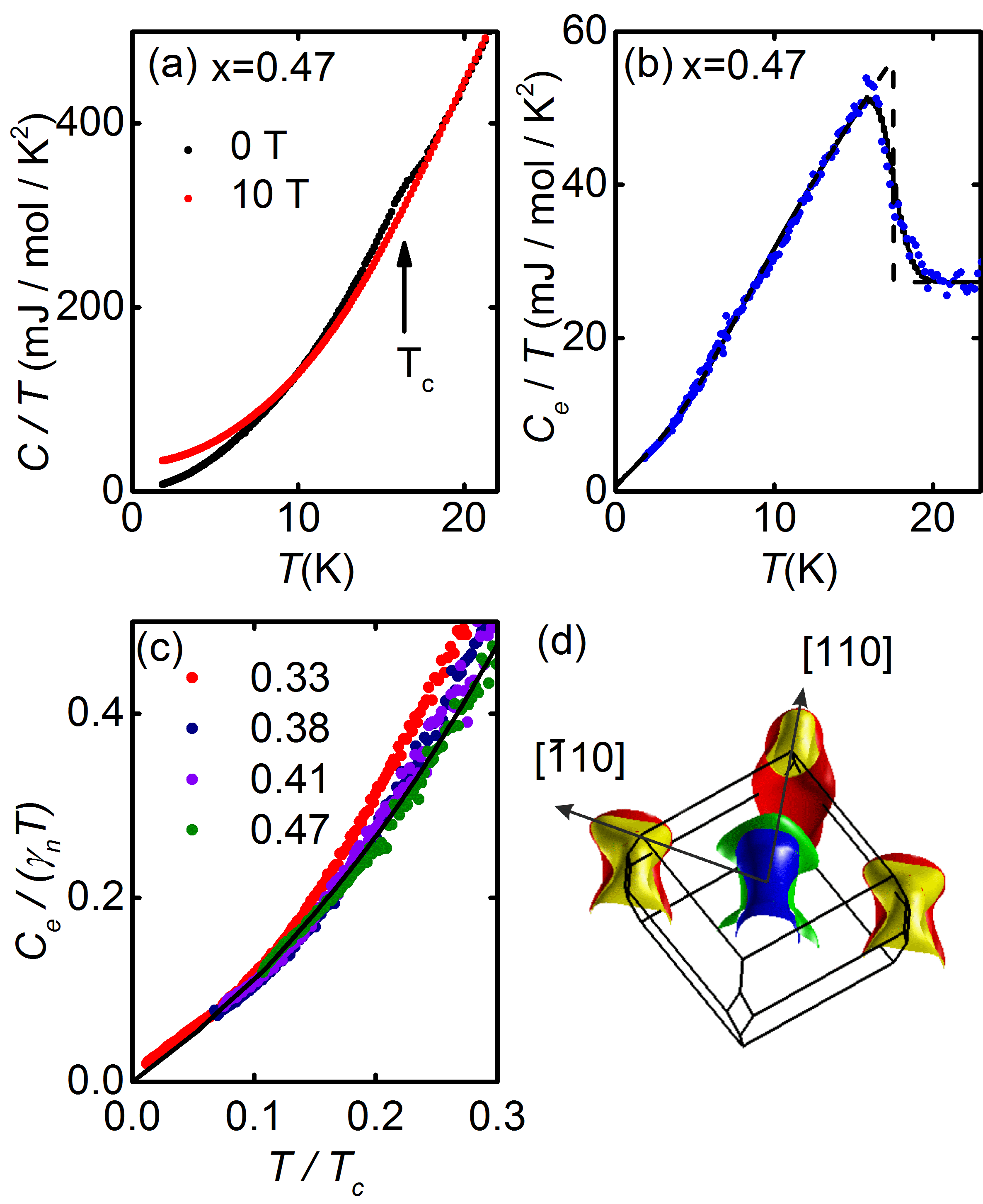}
\caption{(color online).(a) Specific heat versus temperature in 0\,T and 10\,T ($H\|c$) for BaFe$_2$(As$_{1-x}$P$_{x}$)$_2$ with $x=0.47$ ($T_c=17.5$\,K). (b) Electronic
heat capacity in the superconducting state. The solid line is a fit to a nodal structure as described in the text convoluted with a Gaussian spread of transition temperatures $\delta T_c/T_c=0.03$.  The dashed line is
the fit without the convolution. (c) Specific heat versus temperature divided by  $\gamma T_c$ for several
values of $x$.  The line is fit to $x=0.47$ data as shown in (b). (d) Schematic of the Fermi surface showing the electron and hole sheets at the zone corner and center respectively.
}
\label{Fig:Tdep}
\end{figure}

Single crystals of BaFe$_2$(As$_{1-x}$P$_{x}$)$_2$ were grown using a self-flux method \cite{kasahara2010evolution} and
the phosphorous content $x$ was measured using energy dispersive x-ray analysis(EDX) with an accuracy of $\pm 0.02$ in
$x$. The specific heat was measured using a custom built calorimeter \cite{taylor2007specific,malone2010location} run
in two modes: a relaxation mode to measure the temperature dependence at fixed field and an AC mode to measure the
field angle dependence at fixed temperature.  After each rotation of the field, the sample was heated above $T_c$ and
field cooled to minimize any effects of flux pinning.  Our technique allowed us to measure small single crystals; the
samples measured here had masses between 50 and 300 $\mu$g.  The absolute calibration of the calorimeter was checked
with a high purity Ag sample and deviations from standard data were below 1\% in fields up to 14\,T.

\begin{figure}
\includegraphics[width=80mm]{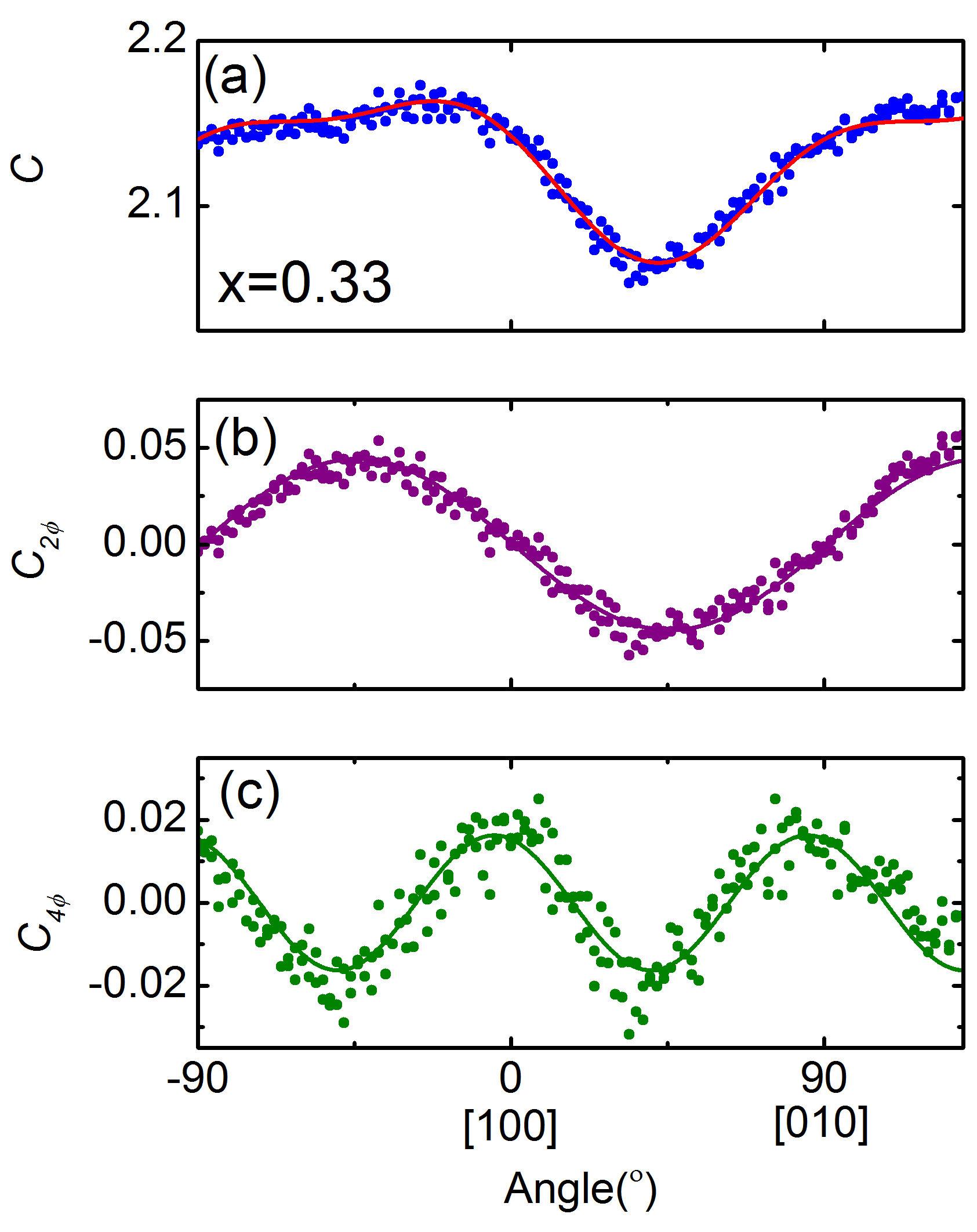}
\caption{(color online).(a) Specific heat of BaFe$_2$(As$_{1-x}$P$_{x}$)$_2$ ($x=0.33$, $T_c=28.8$\,K) as a function of angle as the field is rotated between the \emph{a} and
\emph{b} axis.  The line is a fit to Eq.\ \ref{Eq:rotc}.  (b) and (c) show the same data with the fitted $C_{4\phi}$ and $C_{2\phi}$ terms respectively plus the constant $C_0$ subtracted.
The units of $C$ in all panels are mJ mol$^{-1}$K$^{-1}$.
} \label{Fig:AngleDepC-twofour}
\end{figure}

We begin by discussing the temperature dependence of the specific heat of a sample of BaFe$_2$(As$_{1-x}$P$_{x}$)$_2$
with $x=0.47$. This particular composition was selected because $H_{c2}$ can be exceeded easily with the available
magnetic field. Fig.\ \ref{Fig:Tdep}(a) shows the specific heat of this sample measured in 0\,T and 10\,T.  A
superconducting transition is observed at 17.5\,K in 0\,T which is completely suppressed with a field of 10\,T. To
obtain the electronic contribution to the heat capacity, the 10\,T data is assumed to consist of a field independent
phonon term plus a linear electronic term $\gamma_n T$.  The phonon contribution is then subtracted from the 0\,T data
to obtain the electronic contribution to the specific heat $C_e$ in the superconducting state (Fig.\
\ref{Fig:Tdep}(b)). We have verified that $\int_0^{T_c}C_e/TdT$ is equal to $\gamma_n T_c$ at $T_c$ within experimental
error confirming that the $C_e$ has been accurately isolated. We find that $C_e/T$ varies linearly with $T$ at low
temperature and is well fitted by a nodal gap model (described in more detail later), confirming the presence of gap
nodes at this composition. For higher $T_c$ samples our available magnetic field ($\leq 14$\,T) is insufficient to
completely suppress superconductivity, however, by assuming that the phonon term does not change appreciably with $x$
 we can estimate $C_e$ for all $x$. In Fig.\ \ref{Fig:Tdep}(c) we show the low temperature specific heat
for several different values of $x$ with the same $x=0.47$ phonon term subtracted. The data are shown on normalized
axes ($C_e/\gamma_nT$ versus $T/T_c$), where $\gamma_n$ was determined from the height of the superconducting anomaly
as in Ref.\ \cite{Walmsley2013}.  It can be seen that they all tend towards the same limiting behavior as
$T/T_c\rightarrow 0$ where differences in phonon terms become negligible. This is consistent with Ref.\
\cite{hashimoto2012sharp} where it was found that the size of the linear term in $\lambda(T)/\lambda(T=0)$ was also
independent of $x$.   As the size of linear terms in both $C_e/(\gamma_n T)$ and $\lambda(T)/\lambda(T=0)$ depend on
the number and structure of the gap nodes, these results indicate that the gap structure does not change appreciably
with $x$ in BaFe$_2$(As$_{1-x}$P$_{x}$)$_2$.

We now turn to the question of the direction of the nodes in the superconducting gap. For this study a sample close to
optimal composition ($x=0.33$, $T_c$=28.8\,K) was selected so that we can more easily reach the low $T/T_c$ and
$H/H_{c2}$ limits. Fig.\ \ref{Fig:AngleDepC-twofour} shows the specific heat as a function of field angle at a fixed
temperature as the field is rotated between the \emph{a} and \emph{b} axis.   The data are well fitted by a constant
and two terms which have two fold and four fold symmetry
\begin{equation}\label{Eq:rotc}
 C(\phi)=C_0+C_{2\phi}+C_{4\phi}
\end{equation}
Here $\phi$ is the in-plane angle measured relative to the $a$-axis ([100] direction) and
$C_{n\phi}=c_{n}\cos(n\phi+\delta_n)$. At $T=0.5$\,K and $\mu_0 H = 3$\,T (Fig.\ \ref{Fig:AngleDepC-twofour}) the
twofold component ($C_{2\phi}$) has a peak to peak amplitude of $\sim 5$\,\% of the total $C$, and the fourfold term
$c_4 \simeq 0.5 c_2$.  The phases $\delta_2=89\pm 1^\circ$ and $\delta_4=19\pm 2^\circ$, correspond to the $c_2$ being
zero  at $+1^\circ$ and $c_4$ being maximum at $-5^\circ$ off the $a$-axis. The sample alignment with the field axis is
accurate to $\pm 3^\circ$ so within error both $c_2$ and $c_4$ are approximately aligned with the crystal axes.

\begin{figure}
\includegraphics[width=80mm]{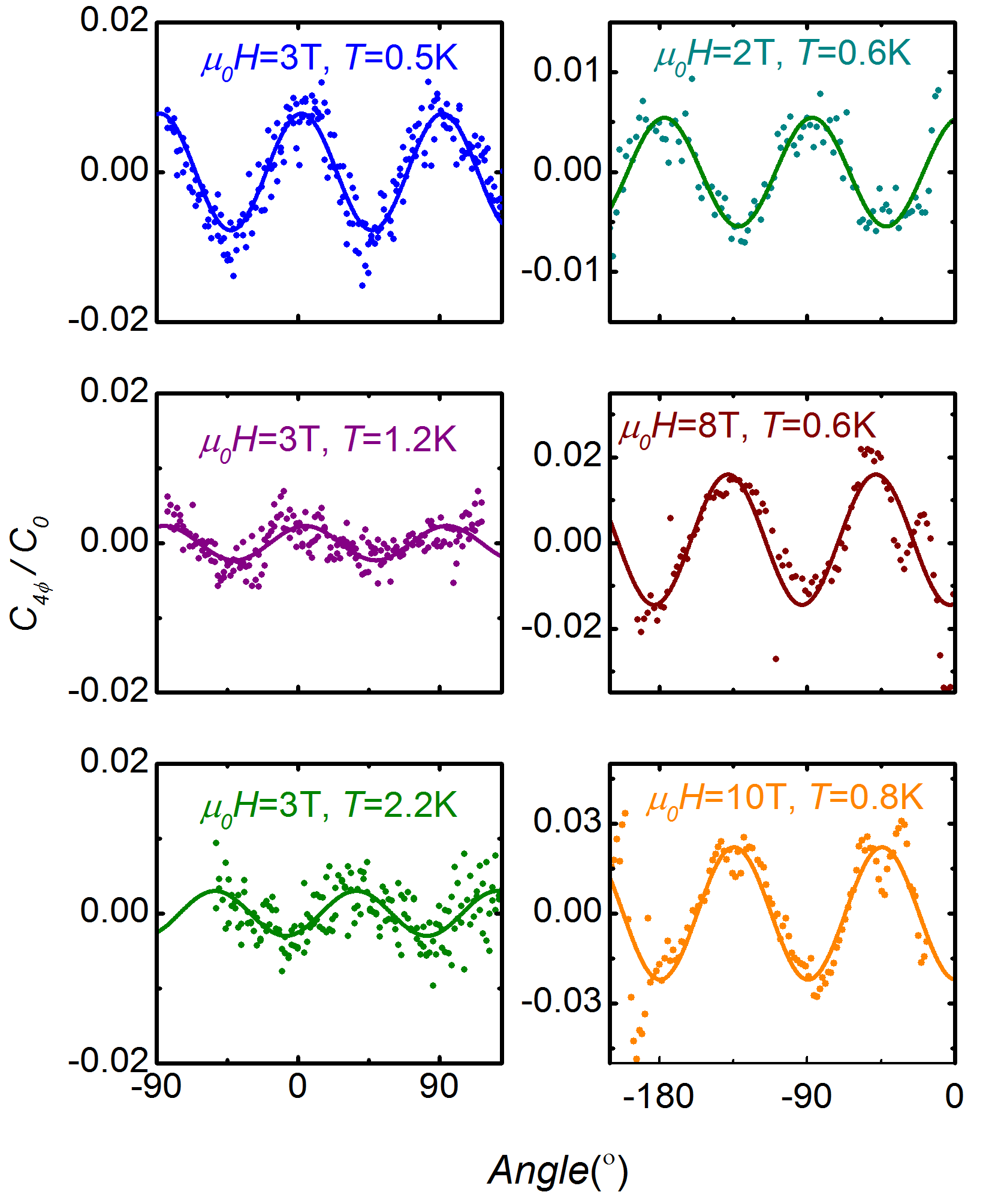}
\caption{(color online). Fourfold component of the same sample as in Fig.\ \ref{Fig:AngleDepC-twofour} at several fields and temperatures.}
\label{Fig:AngleDepC-four-RawTB}
\end{figure}

Theory suggests that there are two mechanisms which can give rise to a field-angle dependence of $C$. First there is a
contribution from the anisotropy of the Fermi velocity $v_F$ and a second from the structure of $\Delta$.  The former
is expected to have a weak dependence on $T$ and $H$ whereas the latter has a much stronger dependence. We find that
for BaFe$_2$(As$_{1-x}$P$_{x}$)$_2$ the twofold component does not depend strongly on $T$ or $H$ \cite{supp} and
therefore likely arises from an anisotropy in $v_F$, similar to the two fold field angular dependence of the normal
state magnetic torque \cite{kasahara2012electronic}.  In Fig.\ \ref{Fig:AngleDepC-four-RawTB} we show the oscillation
data at several different values of $T$ and $B$ with the two fold term subtracted.  We see that the fourfold term
varies strongly as a function of $T$ and $H$ and is therefore likely to originate from the gap anisotropy. Theory
suggests that for a nodal superconductor in the low $T$ and $B$ limit there is a minima in $C$ whenever the field is
applied parallel to a node \cite{vekhter1999anisotropic,vorontsov2010nodes}. This effect has been observed in several
nodal superconductors \cite{aoki2004field,malone2010location}.  So as $c_4$ is minimum when $H$ is aligned
approximately along the [110] directions, we conclude that there are quasi-two-dimensional vertical line nodes located
along these directions. This result is consistent with angle dependent thermal conductivity measurements
\cite{yamashita2011nodal} which showed a four fold term with minima in the same directions as $C$ and also a similarly
strong twofold component.

\begin{figure}
\includegraphics[width=\linewidth]{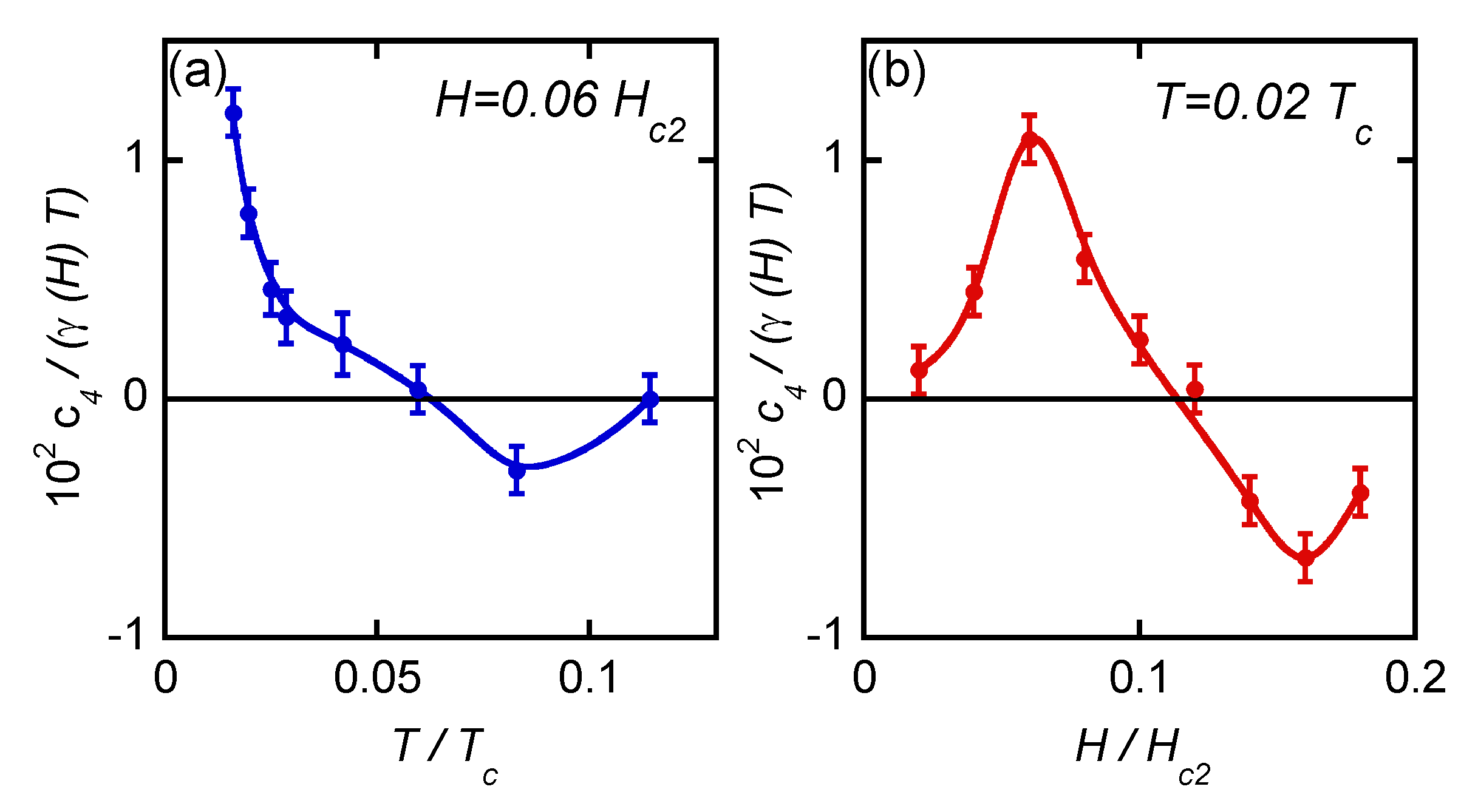}
\caption{(Color Online) Amplitude of the fourfold component of the the specific heat $c_4$, of the same sample as in Fig.\ \ref{Fig:AngleDepC-twofour} as a function of (a)
temperature and (b) field.  $T_c=28.8$\,K and $\mu_0H_{c2}=50$\,T.  Lines are guides to the eye.}
\label{Fig:FourFold-TB}
\end{figure}

Theoretical studies \cite{vekhter1999anisotropic,vorontsov2006nodal} have shown that as the temperature is raised from
zero and the quasiparticle states near the nodes are thermally occupied, the size of the nodal contribution $c_4$
decreases and eventually crosses zero and becomes inverted, so there is then a maximum in $C$ when $H$ is parallel to a
node. Also, as the field increases from zero at low $T$, $c_4$ is expected to increase, reach a maximum then again
cross zero producing a negative $c_4$. In Fig.\ \ref{Fig:FourFold-TB} we show the $T$ and $H$ dependence of $c_4$ in
our sample which displays both of these effects.  $c_4$ decreases sharply as a function of temperature,  crosses zero
at $ T/T_c \simeq 0.05$, and shows a inversion at  $T/T_c \simeq 0.08$.  Following the trend to high temperature
becomes significantly more difficult because of the strongly increasing phonon background.  The field dependence of
$c_4$ displays a maximum at around 3\,T which is $H/H_{c2} \simeq 0.05 $, and then crosses zero at $H/H_{c2}\simeq
0.11$ and becomes inverted and again reaches a negative maximum at $H/H_{c2}\simeq 0.16$.  This behavior is very
similar to that observed in the $d$-wave heavy fermion superconductor CeColn$_5$ where the change in sign of $c_4$
occurs at almost the same reduced field as here \cite{an2010sign}.

Although the angle dependent measurements unambiguously identify the direction of the line-nodes, these measurements
are less clear regarding on which Fermi surface sheet(s) the nodes are located.  In principle, such information might
be extracted from the amplitude of the oscillations and the details of the $T$ and $H$ dependence, although this would
require extensive modelling and may not be robust to the necessary simplifying assumptions of the theoretical model. So
in order to answer this important remaining question we analyze in more detail the temperature and field dependence of
$C$ at fixed angle $H\|c$. For this we use the data for the $x=0.47$ sample were we are most confident of the phonon
background subtraction.

We begin by fitting the temperature dependent data to a simple model, where the gap has a fourfold line node structure
on both the electron (e) and hole (h) sheets,
\begin{equation}
  \Delta_{e,h}=\Delta_0\cos(2\phi).
\label{Eq:gapfun}
\end{equation}
As shown in Fig.\ \ref{Fig:Tdep}b, this model \cite{supp,taylor2007specific} provides an excellent fit of the data with
the single fitting parameter $\Delta_0=2.1\pm 0.1 k_BT_c$ ($\gamma_n=27.7$\,mJ/mol/K$^2$ is fixed by the normal state
value). This gap structure and value of $\Delta_0$ is consistent with $\lambda(T)$ data for the same system. Within the
same gap model, at low temperature $\Delta\lambda(T)/\lambda(0)=(\ln(2) k_B T)/\Delta_0$, giving
$(T_cd\lambda/dT)/\lambda(0)\simeq 0.33$ which is close to the experimental, approximately $x$ independent value of
$0.4\pm0.1$ \cite{hashimoto2012sharp}. This value of $\Delta_0$ is close to that expected from the single band $d$-wave
model in the weak-coupling limit $\Delta_0=2.1 k_BT_c$ \cite{WonMaki1994} although our measurements are not sensitive
to the phase of the $\Delta$ so we can not distinguish between $A_{1g}$ and $B_{1g}$ gap symmetry.  The disorder
dependence of $\lambda(T)$ suggests that the symmetry is $A_{1g}$
 \cite{Mizukami14}.

Although, this form of the gap function provides an excellent description of the data with a single parameter, the fit
alone cannot exclude more complicated multigap forms. Gap functions where there are line or loop nodes on either the
hole or electron sheets but not on both have been proposed \cite{Graser2010,Suzuki11,yamashita2011nodal}. In general,
line nodes will always give a $T^2$ contribution to $C$ for $T\ll T_c$ the coefficient of which ($\alpha$) will depend
on the number of nodes, the gap slope $\eta=d\Delta/d\phi$ and $v_F$ at the node.  So reducing the number of nodes (for
example having them on the electron sheets only) will reduce $\alpha$ but this can be compensated for by reducing
$\eta$. The jump at $T_c$ ($\Delta C/(\gamma_n T_c)$) on the other hand, will depend on the Fermi surface averaged
value of $(d\Delta^2/dT)/v_F$. So in general, the essential parameters of any possible gap function, with respect to
fitting the whole temperature dependence of the specific heat, are the number of nodes divided by the gap slope $\eta$
and the average value of the gap over the whole Fermi surface.  Clearly, there are many possible solutions which would
also provide a reasonable fit to the data in Fig.\ \ref{Fig:Tdep}(b) (see Ref.\ \cite{supp} for further details).
However, all of these have considerably more structure in $\Delta(\bm{k})$ than Eq.\ \ref{Eq:gapfun}.


\begin{figure}
\includegraphics[width=80mm]{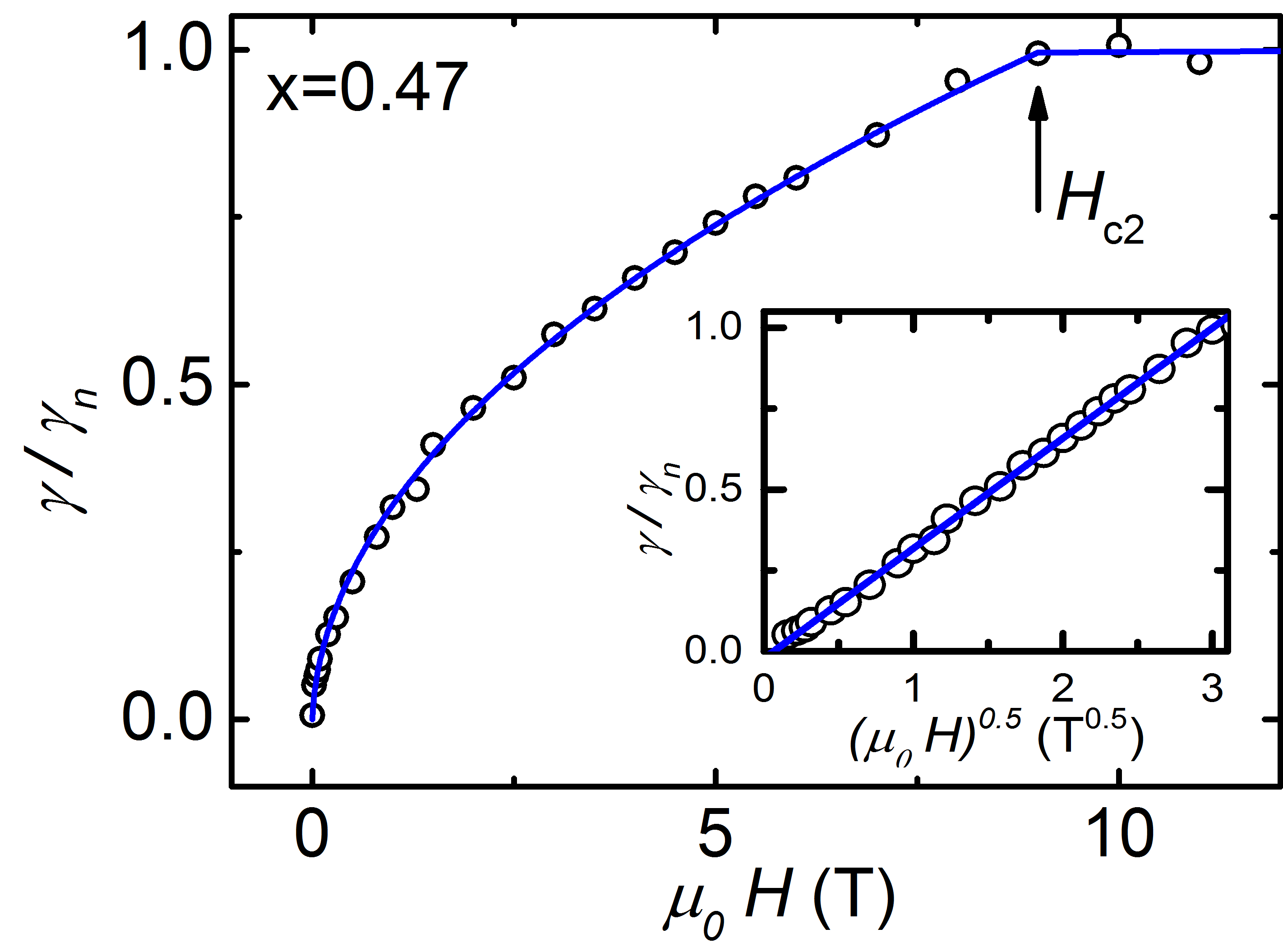}
\caption{(color online). Field dependence of the Sommerfeld coefficient $\gamma$ of BaFe$_2$(As$_{1-x}$P$_{x}$)$_2$ with $x=0.47$.  The line shows a fit
a fit to $\gamma(H)=AH^{\frac{1}{2}}$ below $H_{c2}$ and a constant above this. Inset: The same data plotted versus $H^{\frac{1}{2}}$. }
\label{Fig:RootH}
\end{figure}

To distinguish further between these single or multigap scenarios we turn to the field dependence of the specific heat.
In a nodal superconductor, in the clean limit, at $T=0$ and $0<H\ll H_{c2}$ we should expect that $C_e\sim \gamma T$,
with $\gamma \sim H^{\frac{1}{2}}$. This was first observed in the cuprate superconductor YBa$_2$Cu$_3$O$_{7-\delta}$
\cite{Moler1994} but has now been seen in many different materials.  At higher $H$ we might expect some departure from
this simple form but nevertheless we would expect $\gamma(H)$ to evolve smoothly.  On the other hand, in the multigap
case, we would expect significant structure in $\gamma(H)$.  For example, if there is a much smaller gap on one  sheet
relative to another we would expect $\gamma$ to increase strongly below a field scale set by the small gap and then
more slowly thereafter. This behavior has been experimentally observed in both fully gapped and nodal multigap
superconductors such as MgB$_2$ \cite{Bouquet2002}, KFe$_2$As$_2$ \cite{Hardy2013} and Sr$_2$RuO$_4$
\cite{Deguchi2004}. Also, if there was a large constant gap on one sheet and a nodal gap on the another, we would
expect at low fields a  $H^{\frac{1}{2}}$ contribution from the nodal part and a much smaller linear $H$ contribution
from the fully gapped part \cite{Wang2011}. In this case, as $H$ approaches $H_{c2}$, $\gamma(H)$ would necessarily
have to rise rapidly towards $\gamma_n$.  So the form of $\gamma(H)$ is a further constraint on possible gap
structures.

To extract $\gamma(H)$ for our sample of  BaFe$_2$(As$_{1-x}$P$_{x}$)$_2$ ($x$=0.47) we measure the $T$ dependence at
several fixed fields and fit $C/T$ to a second order polynomial with the phonon term fixed to the high field value. As
shown in Fig.\ \ref{Fig:RootH} we find that $\gamma(H)$ is well described by the nodal model; $\gamma(H)=\beta \gamma_n
(H/H_{c2})^\frac{1}{2}$ from low field all the way up to $H_{c2}$, with $\beta=1.05\pm 0.02$ very close to unity. There
is no additional structure in $\gamma(H)$, such as upturns close to $H_{c2}$, plateaus or other features. This strongly
indicates that there are no fully gapped sheets of Fermi surface.  It suggests that the gap structure is simple and
evolves smoothly with field. We find very similar behavior of $\gamma(H)$ for $x=0.41$ \cite{supp}.


Although our data are most simply explained by the above single nodal gap model, we cannot rule out the possibility of
there being horizontal rather than vertical nodes on some of the Fermi surface sheets.  Also, loop nodes, provided that
the $\phi$ separation of the loop is small, so they approximate a single line node and therefore do not generate higher
harmonics in $C(H,\phi)$, may also be consistent with our data. More detailed theoretical modelling of $C(T,H,\phi)$
with realistic Fermi surface parameters would help to resolve this.  The direction of the nodes is consistent with the
gap anisotropy observed by ARPES measurements \cite{Yoshida14} on one of the electron sheets, however the lack of
observation of nodes on the hole sheets in the same study (and Ref. \cite{shimojima2011orbital}) is inconsistent with
our $\gamma(H)$ results. The results might be reconciled if the nodal hole sheet gap is more sensitive to disorder or
different at the sample surface. Indeed horizontal nodes on one of the hole sheets were reported by a different ARPES
study \cite{zhang2012nodal}.

In summary, our measurement of the specific heat of BaFe$_2$(As$_{1-x}$P$_{x}$)$_2$ show that gap nodes exist along the
[110] (Fe-Fe bond) direction.  The data further suggest that nodes exist on all the Fermi surface sheets. The
quantitative agreement of the low temperature specific heat and magnetic penetration depth for values of $x$ across the
superconducting part of the phase diagram suggest that this gap function does not vary significantly with $x$.

We thank P. Hirschfeld and I. Vekhter for useful discussions.  This work was supported by the Engineering and Physical
Sciences Research Council (Grant No. EP/H025855/1) and Topological Quantum Phenomena" (No. 25103713) Grant-in Aid for
Scientific Research on Innovative Areas from MEXT, and KAKENHI from JSPS.


\newpage

{\large\bf \noindent Supplementary Information}

\subsection{Temperature and field dependence of twofold term}

The temperature and field dependence of the amplitude of the twofold term $c_{2}$ extracted from the specific heat as a
function of field angle is plotted in Figure S1. The variation of $c_{2}$ in the range measured is small relative to
that of $c_4$ (see main text).
\begin{figure}[h]\label{FigS1}
\includegraphics[width=80mm]{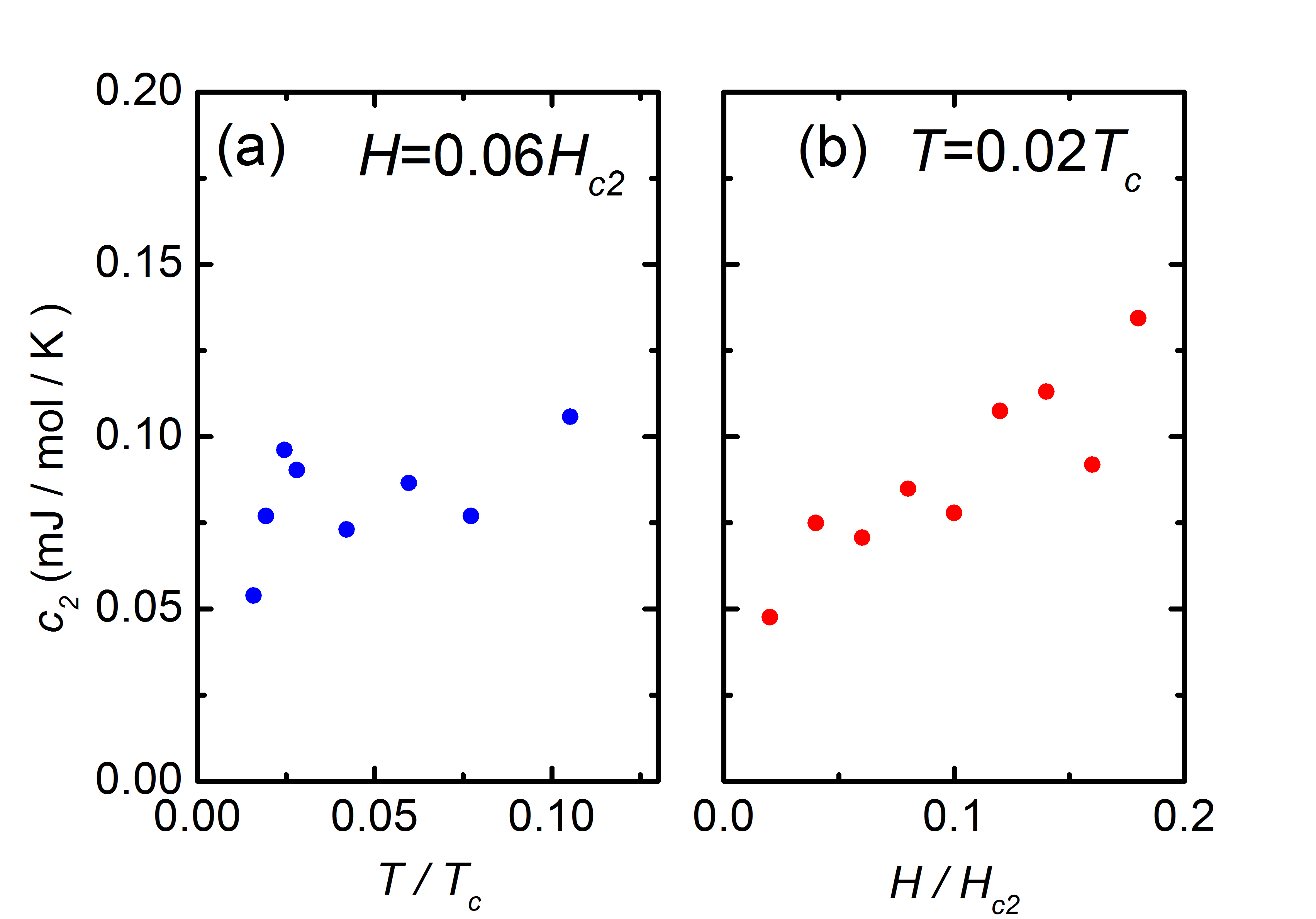}
\caption{Amplitude of the twofold component $c_2$ of the specific heat as a function of (a) temperature and (b) field.}
\end{figure}

\subsection{Field dependence of $\gamma$}

To extract the field dependence of $\gamma$, the specific heat versus temperature is measured in multiple fields, using
a carefully field calibrated calorimeter.  The data are fitted to the form $\frac{C}{T}=\gamma+AT+B T^2$, where the
phonon term $B=1.1$\,mJ/mol/K$^4$ is fixed independent of field.  Figure S3 shows the data with the fits for several
applied fields.
\begin{figure}
\includegraphics[width=80mm]{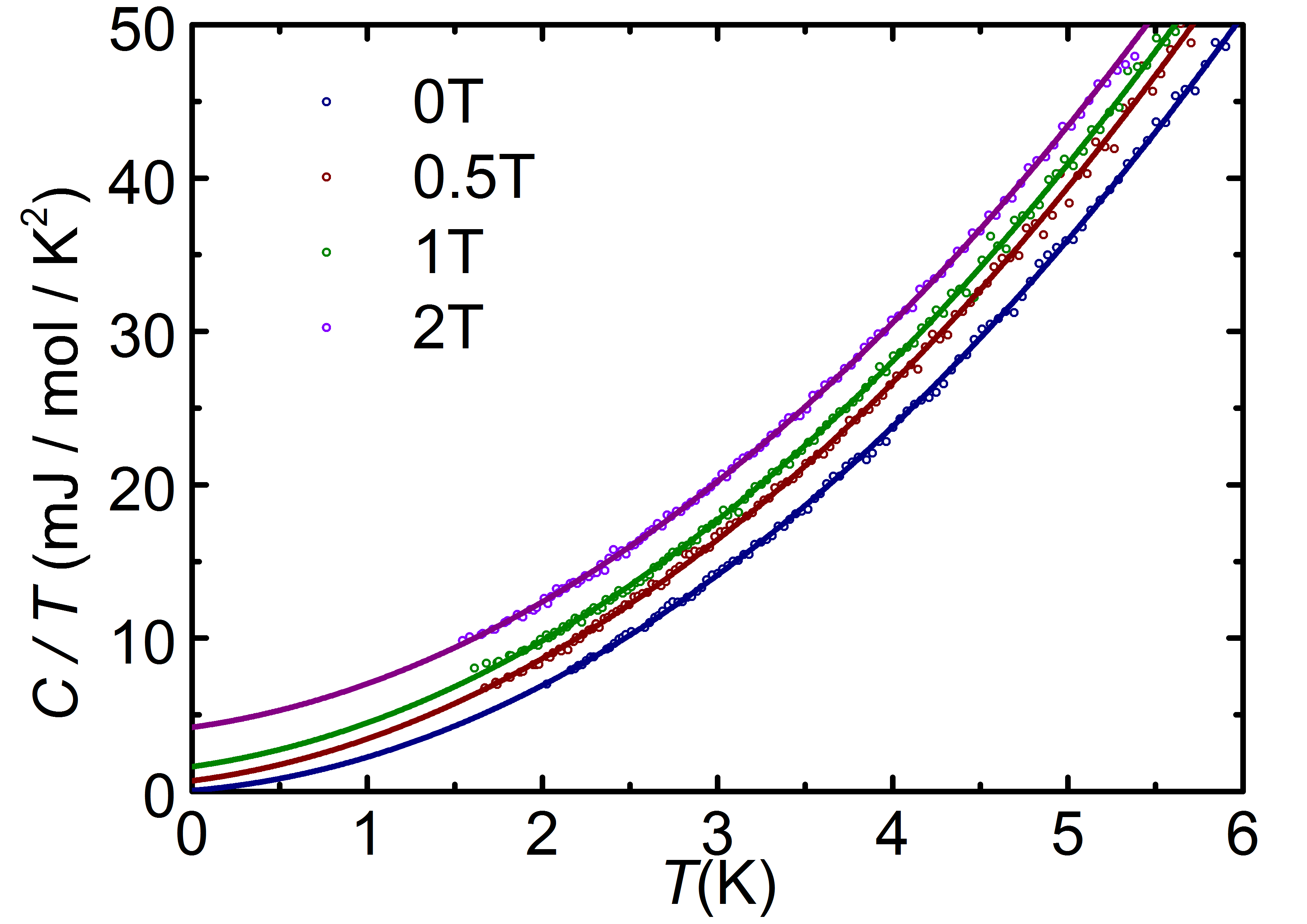}
\caption{ Specific heat of BaFe$_2$(As$_{1-x}$P$_{x}$)$_2$ with $x=0.47$ as a function of temperature in multiple fields.  Lines are fits used to extract $\gamma(H)$. }
\end{figure}

\begin{figure}
\includegraphics[width=80mm]{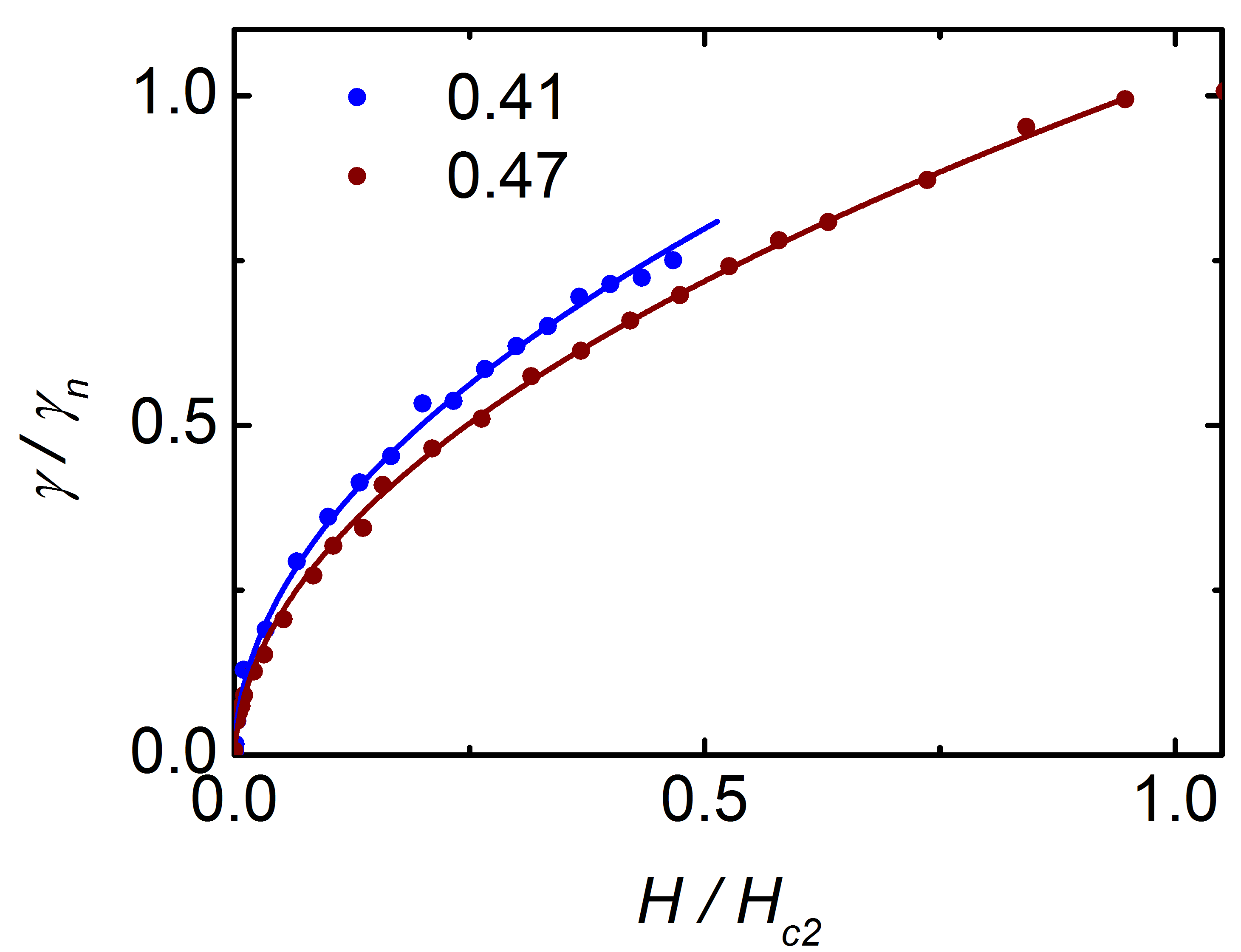}
\caption{Field dependence of $\gamma(H)$ for two samples of BaFe$_2$(As$_{1-x}$P$_{x}$)$_2$, $x=0.41$ and $x=0.47$.}
\end{figure}

Figure S4 shows $\gamma(H)$ for the 0.47 and 0.41 samples. For the $x=0.41$ sample, $\gamma$ was extracted from the
jump at $T_c$ to be 38\,mJ/mol/K$^2$ and $H_{c2}$ was determined to be  30\,T by comparison to samples with the same
$x$ and $T_c$ \cite{PutzkeAnom}. The $\gamma(H)$ data are fitted to
$\frac{\gamma}{\gamma_n}=\beta(\frac{H}{H_{c2}})^{1/2}$, the fitted values of $\beta$ are 1.14$\pm0.1$ for x=0.41 where
the error primarily arises from the uncertainty on $H_{c2}$ in this sample and 1.05$\pm0.02$ for $x=0.47$.  Within
error the size of $\beta$ in the two samples is equal.

\clearpage
\newpage

\subsection{Fitting temperature dependence of specific heat}

\begin{figure*}
\label{Fig5}
\includegraphics[width=160mm]{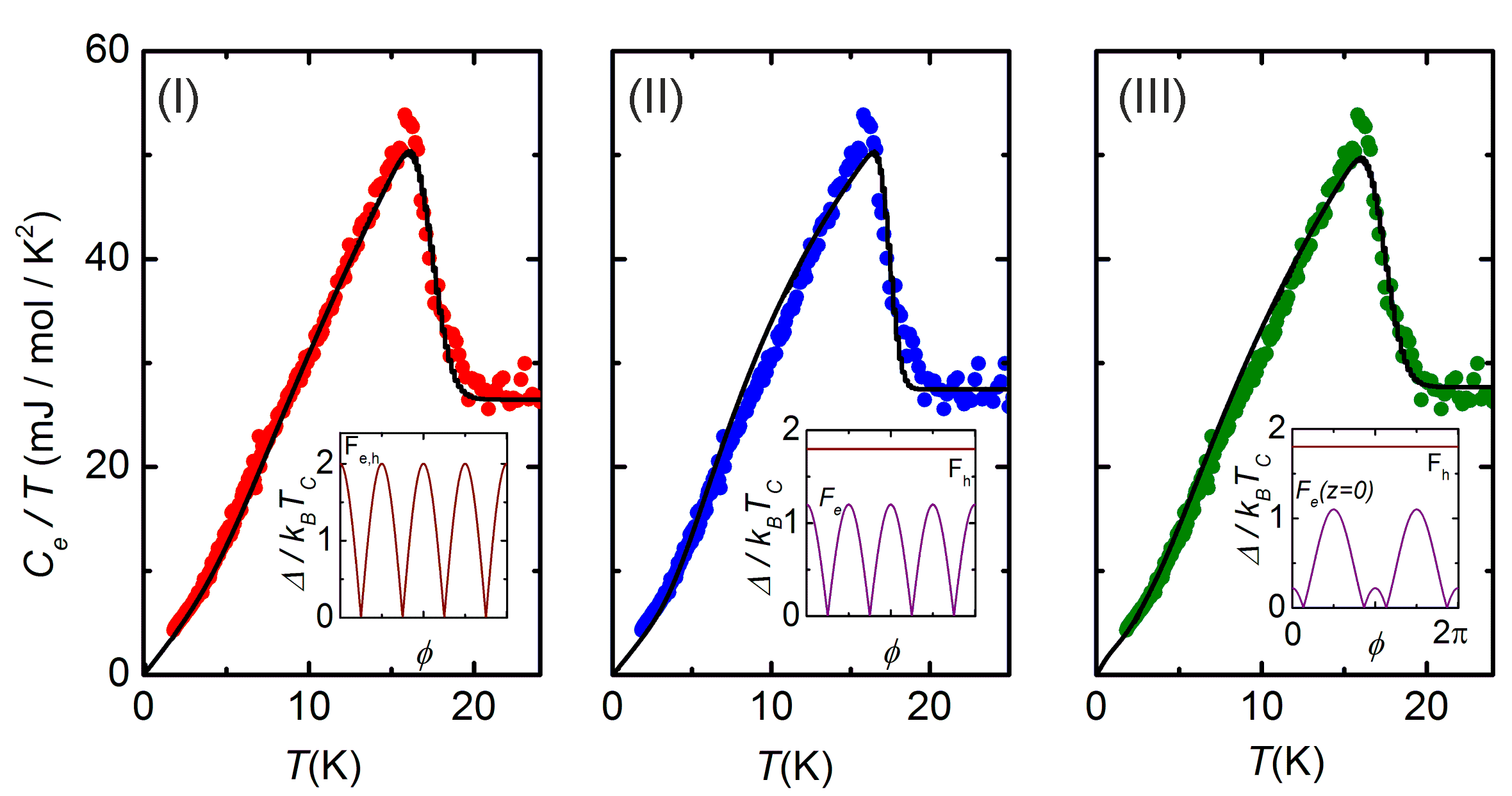}
\caption{Electronic specific in the superconducting state of $x=0.47$ sample with three different fits.  Case (I) nodes
on all Fermi surface sheets. Case (II) vertical nodes on the electron sheets and isotropic gap on the hole sheets. Case
(III) loop nodes on the electron sheets and isotropic gap on the hole sheets.  The insets are the fitted gap structures
on each sheet.}
\end{figure*}

We calculated the temperature dependence of the specific heat for a generalized superconducting energy gap function
$\Delta_{i}=F_{i}(\phi,z)$ (where $i$ represents each Fermi surface sheet and $\phi, z$ are cylindrical coordinates),
using the following equation for the entropy
\begin{equation}
S_i=-\frac{k_B}{2\pi^3\hbar}\int_0^\infty \int\frac{1}{v_F(\mathbf{k})}([1-f]\ln[1-f]+f\ln f)dS_fd\epsilon.
\end{equation}
Here $f$ is the fermi function $f=[\exp(E/(k_BT))+1]^{-1}$, $E$ is the quasiparticle energy
$E^2=(\epsilon^2+\Delta(\phi,z)^2)$, $\epsilon$ is the single particle energy, $v_F(\mathbf{k})$ is the normal state
Fermi velocity and the integral $dS_F$ is over each sheet of Fermi surface. The total heat capacity is then calculated
using $C=T\sum\frac{\delta S_i}{\delta T}$. We found that for the gap structure considered here,  the anisotropy of
$v_F(\mathbf{k})$ (which we obtained from DFT band structure calculations) makes little difference to the specific heat
fits and therefore we adopt a simpler form for the entropy where we approximate each Fermi sheet as a cylinder and
weight the contribution of each sheet according to its relative contribution to the total density of states $N$. To
obtain the relative weights of $N$ we use the Fermi surface of BaFe$_2$P$_2$, which gives : 20\,\% (outer electron),
20\,\% (inner electron), 23\,\% (inner hole) and 37\,\% (outer hole).   It has been shown experimentally
 that the mass enhancements on each sheet are relatively uniform  for all $x$ \cite{Arnold2012,walmsley2013}. The total
magnitude of $N$ is set by $\gamma_n$ which is experimentally determined above $T_c$. Finally, the fit is convoluted
with a Gaussian spread of transition temperatures to describe the rounding near $T_c$. For the $x=0.47$ sample this
spread is found to be $\delta T_c/T_c=0.03$.

In a nodal superconductor, the amplitude of the low temperature linear behavior is determined by the slope of the gap
at the node $\eta=\frac{d\Delta}{d\phi}$ and the number of nodes. The simplest nodal function (Case (I)) has
$F_{e,h}=\Delta_0\cos(2\phi)$, and as the gap has the same structure on each sheet the relative weighting is redundant.
This model fits the data very well over the entire temperature range as shown in Figure S2 and main text Figure 1. As
discussed in the main text it is also possible to fit the data with a number of other different gap structures.  To
illustrate this point, Figures S2(b,c) show the data fitted to (Case (II)) a gap structure which has vertical nodes on
the electron sheets and an isotropic gap on the hole sheets, and (Case (III))  a loop node structure (as suggested in
Ref.\ \cite{yamashita2011nodals}) where the gap on the electron sheets follows,
$F_e=\Delta_e(1-|0.6\cos(z)|-0.6\cos(z)\cos(2\phi))$) and the gap on the hole sheets is isotropic. The gap structures
can be visualized in the insets to Figure S2.  Comparing Case (I) to Case (II), it can be seen that  in (II) the
maximum gap on the electron sheets has decreased thus decreasing $\eta$ and compensating for the lack of nodes on the
hole sheets with respect to fitting the low temperature linear term in $C/T$. In addition, the isotropic gap on the
hole sheet in (II) is increased relative to the average gap in (I), so that the size of the jump in $C$ at $T_c$ is
reproduced.  This illustrates the general principles outlined in the main text.  For both (II) and (III) the fit is
slightly worse than for model (I) even though there are more parameters, however, this alone is not sufficient to
distinguish between the models. For example, the fit can be improved by allowing he exact form of the gap function in
the loop node model to vary.  However, as argued in the main text, the additional structure of the gaps in cases (II)
and (III) compared to (I) would inevitable lead to additional structure in $\gamma(H)$ which was not observed.  We
therefore conclude that the thermodynamic data point to there being a simple nodal structure on all sheets.  The
structure does not have to be exactly the same as (I), for example we cannot rule out there being horizontal nodes on
one of the sheets, but there must be nodes on all sheets to reproduce the observed $\gamma(H)$.

\end{document}